\documentclass[twocolumn,showpacs,preprintnumbers,amsmath,amssymb]{revtex4}

\usepackage{graphicx}
\usepackage{dcolumn}
\usepackage{bm}

\begin{document}

\preprint{}

\title{Phase Transitions in Neutron Stars and Gravitational Wave Emission.}

\author{G. F. Marranghello}
\email{gfm@if.ufrgs.br}
 \altaffiliation{}
\author{Cesar A. Z. Vasconcellos}%
 \email{cesarzen@if.ufrgs.br}
\affiliation{Instituto de F\'{i}sica, Universidade Federal do Rio Grande
 do Sul, Porto Alegre, Brazil.
}%

\author{J. A. de Freitas Pacheco}
\email{pacheco@obs-nice.fr}
\affiliation{Observatoire de la C\^ote d`Azur, Nice, France.}%

\date{\today}

\begin{abstract}
We review the detectability of gravitational waves  generated by oscillations
excited during a phase transition from hadronic matter to deconfined quark-gluon matter in the core of a neutron
star. Neutron star properties were computed using a Boguta and Bodmer's based model and the MIT bag model. The
maximum energy available to excite mechanical oscillations into the star is estimated by energy difference between
the configurations with and without a quark-gluon matter core. On basis of the planned sensitivity of present laser
interferometers (VIRGO or LIGO I) and those of the next generation (LIGO II), the maximum volume to be probed by
these experiments is determined. These results are used as an indication of the potential detectability of neutron
stars as sources of gravitational waves. Our results indicate that the maximum distance probed by the detectors of
the first generation is well beyond M31, whereas the second generation detectors will probably see phase transition
events at distances two times longer, but certainly not yet attaining the Virgo cluster.
\end{abstract}

\maketitle

\section {Introduction}

The first generation of large gravitational interferometric detectors as the
French-Italian VIRGO and the American LIGO, should be fully operational within
one or two years. The best signal-to-noise (S/N) ratio that can be achieved from
these detectors implies the use of matched-filter techniques, that require a
priori the knowledge of the signal waveform. Thus, the identification of
possible sources having a well defined signal is a relevant problem in the
detection strategy.

Neutron stars are certainly one of the most popular potential sources of gravitational waves (GWs), since they can
emit by different mechanisms having a well known waveform. Rotating neutron stars may have a time-varying quadrupole
moment and hence radiate GWs, by either having a triaxial shape or a misalignment between the symmetry and the spin
axes, which produces a precessional motion\cite{ferrari69,zimmerman79,shapiro83}. Moreover, fast rotating
proto-neutron stars may develop different instabilities such as the so-called Chandrasekhar-Friedman-Schutz (CFS)
instability\cite{chadrasekhar70,friedman78}, responsible for the excitation of density waves traveling around the
star in the sense opposite to its rotation, or undergo a transition from axi-symmetric to triaxial shapes through
the dynamical ``bar-mode'' instability\cite{lai95}. All these mechanisms are potentially able to generate large
amounts of energy in the form of GWs.

Nonradial oscillations are also a possible mechanism for neutron stars emit GWs and
this possibility was already discussed in the late sixties\cite{thorne67}. One of the difficulties
with this mechanism concerns the energy source necessary to excite the oscillations. The
elastic energy stored in the crust and released by tectonic activity was recently
considered\cite{pacheco98}, but the maximum available energy is likely to be of the order of
10$^{44-45}$ erg. Thus, even if all this energy could be converted into nonradial
modes, the maximum distance that a signal could be seen by a laser interferometer like
 VIRGO is only about 3.0 kpc\cite{pacheco98}.
This result depends, of course, on the adopted equation of state, which fixes the mode frequencies and damping
timescales. A considerable amount of energy would be available if the neutron star undergo a phase transition in the
core. For instance, if quark deconfinement occurs, then the star will suffer a micro-collapse since the equation of
state of the quark matter is softer than that of the hadronic matter and the new equilibrium configuration will be
more compact, having a larger binding energy. The energy difference will partially cover the cost of the phase
transition and will partially be used to excite mechanical modes, whose energy will be dissipated by different
channels. The physical conditions of matter at densities above the saturation density ($\rho_0 = 0.15 fm^{-3}$)
is badly known and many first or second order phase transitions have been speculated upon. The possibility
of a quark and hadron mixed phase was considered in ref.\cite{glendenning92}, where
the conservation of the electric and the baryonic charges are satisfied not locally but within a volume including 
many droplets  of quarks and of hadrons. This mixed phase does not include local surface and Coulomb
energies of quarks and nuclear matter and, as a consequence of these terms, if the interface tension
between quark and nuclear matter is too large, the mixed phase is not favored energetically \cite{heiselberg,jensen}. In this
case, the neutron star will then have a core of pure quark matter and a mantle of hadrons
surrounding it. In the absence of dynamical models able to follow the evolution of the star during
the phase transition, we assume as in ref.\cite{haensel82}, that    
the structural rearrangement suffered by a ``cold'' star occurs on a dynamical timescale, i.e., of the
order of milliseconds, shorter than the smooth global stellar
readjustment suggested in refs.\cite{glen1,glen2} and the gradual transition expected to occur in hot proto-neutron
stars\cite{baumgarte96}. Radial modes excited by the micro-collapse do not radiate GWs, but an important coupling
with rotation exists\cite{chau67} and, if the star rotates, the oscillations will be damped not only by dissipation
of the mechanical energy into heat but also by the emission of GWs.

In the present work, the detectability of GWs generated by oscillations excited during a phase transition in the
core of a neutron star is reviewed\cite{sotani}. Neutron star properties were computed using a description of the hadronic matter
based on the work of Boguta and Bodmer\cite{boguta77}, including the fundamental baryon octet, the isovector meson
$\varrho$ and lepton degrees of freedom (see references [15] for details). Hybrid models, including a quark-gluon
core, with the {\it same baryonic number} as the pure hadron configuration were also computed and the energy
difference between both states was used as an estimate of the maximum energy available to either excite mechanical
oscillations or to be converted into heat. Then, using the planned sensitivity of present laser interferometers like
VIRGO (or LIGO I) and those of the next generation (LIGO II), the maximum volume of space that can be probed by
these experiments is calculated, since this indicates the potential detectability of these sources. The plan of this
paper is the following: in section 2 the star models are reviewed, in section 3 the gravitational wave emission is
discussed and finally, in section 4, conclusions are given.

\section{The neutron star model}

The present model for the nuclear matter was already discussed by Marranghello et al.\cite{marranghello01}, who have
investigated the effects of a finite temperature in the equation of state. Here, the same description for dense
matter is adopted but only models with T=0 will be considered. Moreover, the coupling constants are slightly
modified with respect to those considered in that work, in order to allow the maximum mass of the configuration to
be compatible with recent results on the binary X-ray system Vela X-1\cite{barziv01}. For the sake of completeness, we
recall here the main points of the model.

The lagrangian density describing the nuclear matter is

\begin{eqnarray}
{\cal L}    &=& \sum\limits_{B}   \bar{\psi}_{B} [( i\gamma_\mu (\partial^\mu- g_{\omega B} \omega^{\mu}) -
(M_B-g_{\sigma B} \sigma)
]\psi_B \nonumber \\ && - \sum\limits_{B}   \bar{\psi}_{B} [
\frac12 g_{\varrho B} \mbox{\boldmath$\tau$} \cdot \mbox{\boldmath$\varrho$}^\mu] \psi_B
+\frac{bM}{3}\sigma^3+\frac{c}{4}\sigma^4 \nonumber \\ &&+\frac12(\partial_\mu \sigma \partial^\mu \sigma   -
{m_\sigma^2} \sigma^2)  - \frac14  \omega_{\mu \nu}  \omega^{\mu \nu}   + \frac12 {m_\omega^2}
 \omega_\mu \omega^\mu  \nonumber \\
&& -   \frac14 \mbox{\boldmath$\varrho$}_{\mu \nu} \cdot \mbox{\boldmath$\varrho$}^{\mu \nu} +  \frac12m_\varrho^2
\mbox{\boldmath$\varrho$}_\mu
 \cdot  \mbox{\boldmath$\varrho$}^\mu \nonumber \\
&&+\sum\limits_{l}   \bar{\psi}_{l} [i \gamma_\mu \partial^\mu   - M_l] \psi_l \,\, .
\end{eqnarray}
This equation represents nuclear matter as composed by a mixture of the fundamental baryon octet (p, n, $\Lambda$,
$\Sigma^+$, $\Sigma^0$, $\Sigma^-$, $\Xi^-$, $\Xi^0$) coupled to three mesons ($\sigma, \omega, \varrho$) and
leptons (for the details see [15]). The scalar and vector coupling constants, g$_{\sigma}$, g$_{\omega}$ and the
coefficients b, c were determined by imposing that the model bulk properties should be able to reproduce the binding
energy E$_b$ (= -16.3 MeV), the compression modulus K (= 240 MeV) and the nucleon effective mass $M^* = M
-g_{\sigma} \bar{\sigma}$ (= 732 MeV) at the saturation density $\rho_0$ (=0.153 fm$^{-3}$). Additionally, the
isovector coupling constant g$_{\varrho}$ is determined from the coefficient for the symmetry energy in nuclear
matter, a$_4$ (= 32.5 MeV). We have used the universal hyperon-nucleon coupling 
ratios $\chi_i = g_{Hi}/g_i$, with $i = \sigma, \varrho,
\omega$. The resulting
coefficients used in our computations, are given in table 1. Figure 1 shows the equation of state derived from our
model and in figure 2 it is shown 
the energy density profile inside the star for two
configurations having gravitational masses equal to 1.2 and 1.6 M$_{\odot}$ respectively.

\begin{table}
\caption{\label{tab:table1} }
\begin{ruledtabular}
\begin{tabular}{lcrlc}
$(g_{\sigma}/m_{\sigma})^2$&$(g_{\omega}/m_{\omega})^2$&$(g_{\varrho}/m_{\varrho})^2$&b$(\times 100)$&c$(\times 100)$ \\
\hline
9.927&4.820&4.791&0.8659&-0.2421\\

\end{tabular}
\end{ruledtabular}
\end{table}

\begin{figure}[htb]
\vspace*{10pt} 
\vspace*{1.4truein}             
\vspace*{10pt} \parbox[h]{4.5cm}{ \includegraphics{prd3.ps}}
\vspace{20pt} \caption{Equation of state for hadronic matter (solid line) and
for the quark-gluon matter (dotted line). \label{1}}
\end{figure}

\begin{figure}[htb]
\vspace*{10pt} 
\vspace*{1.4truein}             
\vspace*{10pt} \parbox[h]{4.5cm}{ \includegraphics{prd2.ps}}
\vspace{20pt} \caption{Neutron star energy density-radius relation for M=1.6M$_\odot$ (solid line) and
M=1.2M$_\odot$ (dotted line) on which the quark-gluon core extends up to 8.6 km and 7 km, respectively. \label{3}}
\end{figure}

\begin{figure}[htb]
\vspace*{10pt} 
\vspace*{1.4truein}             
\vspace*{10pt} \parbox[h]{4.5cm}{ \includegraphics{prd4.ps}}
\vspace{20pt} \caption{Neutron star mass as a function of central energy density
for hybrid star with constant pressure transition (solid line), hybrid (dotted
line) and pure hadronis star (dashed line).
configurations \label{2}}
\end{figure}

Hybrid models including a quark-gluon core were also computed. The quark matter was described by the MIT
Lagrangian\cite{grand75} and the physical conditions at the deconfinement transition were estimated from the Gibbs
criteria, namely, by the equality of the chemical potential and pressure of both phases, under conservation of the
baryon number and electrical charge. The physical parameters at the transition point are given in 
table 2, for a bag constant $B^{1/4}=180 MeV$
and a strange quark mass m$_s$ = 150 MeV. The gravitational mass of the star as a function of the central
energy density is shown in figure 3 for the case of a pure hadronic configuration (dashed line), a hybrid
star with a quark core (solid line) and, for comparison, a case where a mixed phase exist in the center,
compute in the same way as ref. \cite{glendenning92}. 

It should be emphasized that our equation of state is
quite steep (${{dlogP}\over{dlog{\varrho}}} \approx 2.6$) near saturation and, as a consequence,
 the deconfinement transition occurs at densities just above the saturation value ($\rho \sim 1.7\rho_0$),
producing hybrid stars with very extended quark-gluon cores. Here we take the opportunity to
reiterate that, from the actual status of theoretical predictions to the EOS,
with so many parameters to be adjusted, even in the quark phase and specially in
the hadron phase, we believe that only qualitative results can be obtained with some
insights on the quantitative results.

\begin{table}[htb]
\begin{center}
\caption{Physical conditions at the  phase transition: energy densities and pressure are given in GeV fm$^{-3}$ }
\vspace{0.5cm}
\begin{tabular}{|lll|}
\hline $\epsilon_H$ & $\epsilon_q$ &P \\ \hline 0.236 & 0.349 & 0.0163\\ \hline
\end{tabular}
\end{center}
\end{table}

The maximum stable mass of a pure hadronic configuration is about M$_{max} \approx$ 2.1 M$_{\odot}$ while for hybrid
stars this limit is reduced to 1.73 M$_{\odot}$. Thus, the present calculations exclude the possibility that Vela
X-1 has a quark core. Notice that our models obey the Seidov criterium\cite{seidov71}, namely, that
the hybrid star
will be stable only if the energy jump across the transition surface satisfies the condition
\begin{equation}
{{\epsilon_q}\over{\epsilon_H}} < {{3}\over{2}}(1 + {{P}\over{\epsilon_H}})
\end{equation}
where  P, e$_q$, e$_H$ are respectively the pressure, the energy density of  quarks and hadrons
at the transition point.

One of the goals of this work is the determination of an upper limit for the energy able to excite the different
oscillation modes of the star, when a modification in its internal structure occurs. In this sense, the simplest
approach is to compute the energy difference between two configurations having the {\it same baryonic number}; the
first constituted of pure hadrons (H), the second having a core of deconfined matter (HQ), and use such a difference
as an indication of the maximum available energy. Table 3 gives the parameters for five models defined by a given
baryonic number\cite{Weinberg}
\begin{equation}
N = \int_0^R 4 \pi r^2 \left[ 1 - \frac{2 G m(r)}{r^2} \right]^{-1/2} \rho_B(r) dr \, ,
\end{equation}
or the baryonic mass of the star ($M_{bar} = N M_B$); the expected gravitational mass of both configurations (pure
hadron and hybrid),
\begin{equation}
M_g = m(R) =  \int_0^R 4 \pi r^2 \epsilon(r) dr \, ;
\end{equation}
expected radii, masses of the quark-gluon core and the energy difference obtained from the passage of configuration
H to HQ, the binding energy.
\begin{equation}
E_g = M_g - M_{bar} \, .
\end{equation}

\begin{table}[htb]
\begin{center}
\caption{Parameters of the stellar models:
 masses are in solar units and
radii in km; H and HQ mean pure hadron and hybrid configurations respectively} \vspace{0.5cm}
\begin{tabular}{|lllllll|}
\hline M$_{bar}$ & M$_g$ (H)&M$_g$ (HQ)&M$_g$ (core)& R (H)& R(HQ)& log (E) erg\\ \hline
1.0749&1.0003&0.9625&0.1150&12.82&12.47&52.83\\ 1.3202&1.2009&1.0953&0.3608&13.04&12.16&53.28\\
1.5724&1.4008&1.2468&0.6013&13.13&11.98&53.44\\ 1.8342&1.6006&1.4164&0.8626&13.01&11.82&53.52\\
2.1045&1.8002&1.6141&1.1766&12.69&11.50&53.52\\ \hline
\end{tabular}
\end{center}
\end{table}

\section{The gravitational wave emission}

The transition from configuration H to HQ may occur through the formation of
a mestastable core, built up by an increasing central density. The increase
in the central density may be a consequence of
a continuous spin-down or other different mechanisms the star could suffer. This transition releases energy, exciting mainly the radial modes of
the star\cite{sotani}. These modes do not emit GWs, unless when coupled with
rotation\cite{chau67}, a situation which will be assumed here.

In order to simplify our analysis, we will consider that most of the mechanical energy
is in the fundamental model. In this case, the gravitational strain amplitude can be
written as
\begin{equation}
h(t) = h_0e^{-(t/{\tau_{gw}}-\imath\omega_{0} t)}
\end{equation}
where h$_0$ is the initial amplitude, $\omega_0$ is the angular frequency of the mode and $\tau_{gw}$ is the
corresponding damping timescale. The initial amplitude is related to the total energy E$_g$ dissipated under the
form of GWs by the relation\cite{pacheco01}
\begin{equation}
h_0 = {{4}\over{\omega_0 r}} {\left[{GE_g}\over{\tau_{gw} c^3} \right]^{1/2}}
\end{equation}
where G is the gravitational constant, c is the velocity of light and r is the distance to
the source.

Relativistic calculations of radial oscillations of neutron star with a quark core were recently performed by Sahu
et al.\cite{sahu01}. However, the relativistic models computed by those authors do not have a surface of
discontinuity  where an energy jump occurs. Instead a mixing region was considered, where the charges (electric and
baryonic) are conserved globally but not locally\cite{glendenning92}. Oscillations of star models including an
abrupt transition between the mantle and the core were considered by
Haensel\cite{haensel89} and Miniutti\cite{miniutti}. However a Newtonian treatment
was adopted and the equation of state used in the calculations does not correspond to any specific nuclear
interaction model. In spite of these simplifications, these hybrid models suggest that {\it rapid} phase
transitions, as that resulting from the formation of a pion condensate, proceed at the rate of strong interactions
and affect substantially the mode frequencies. However, the situation  is quite different for {\it slow} phase
transitions (the present case), where the mode frequencies are quite similar to those of a {\it one-phase}
star\cite{haensel89}. In this case, scaling the results of  \cite{haensel89}, the frequency of the fundamental mode (uncorrected
for gravitational redshift) is given approximately by
\begin{equation}
\nu_0 \approx 63.8 \left[ {{(M/M_{\odot})}\over{R^3}} \right]^{1/2} \,\ \,\ kHz
\end{equation}
where the mass is given in solar units and the radius in km.

Once the transition to quark-gluon matter occurs, the weak interaction processes for the quarks u, d and s
\begin{equation}
u + s \rightarrow d + u
\end{equation}
and
\begin{equation}
d + u \rightarrow u + s
\end{equation}
will take place. Since these reactions are relatively
slow, they are not balanced  while the oscillations last and thus, they dissipate
mechanical energy into
heat\cite{wang84}. According to calculations of reference [21], the dissipation
timescale can be estimated by the relation
\begin{equation}
\tau_d \approx 0.01{\left({{150 MeV}\over{m_s}}\right)^4}\left({{M_{\odot}}\over{M_c}}\right) \,\ s
\end{equation}
where $m_s$ is the mass of the s-quark in MeV and M$_c$ is the mass of
the deconfined core in solar masses. This equation is valid for temperatures in
the range $10^8-10^9K$. On the other hand, according to reference \cite{chau67}
the damping timescale by GW emission is
\begin{equation}
\tau_{gw} = 1.8\left({{M_{\odot}}\over{M}}\right)\left({{P^4_{ms}}\over{R^2}}\right)    \,\  s
\end{equation}
where again the stellar mass is in solar units, the radius is in km and the rotation
period P is in milliseconds.

In a first approximation, the fraction of the mechanical energy which will be
dissipated under the form of GWs  is  $f_g = {{1}\over{(1 + \tau_{gw}/\tau_d)}}$.
Notice that the damping timescale by GW emission depends strongly on the
rotation period. Therefore one should expect that slow rotators will dissipate
mostly of the mechanical energy into heat.  In table 4
is given for each star model the expected frequency of the fundamental mode (corrected
for the gravitational redshift) , the critical rotation period (in ms) for having f$_g$ = 0.50, the
GW damping for this critical period and the quality factor of the
oscillation, Q =$\pi\nu_{0}\tau_{gw}$.

\begin{table}[htb]
\begin{center}
\caption{Oscillation parameters: the damping timescale $\tau_{gw}$ is given for the critical period; maximum
distances for VIRGO (V) and LIGO II (L) are in Mpc}
\vspace{0.5cm}
\begin{tabular}{|llllll|} \hline $\nu_{0}$ &
$P_{crit}$ & $\tau_{gw}$&Q&D$_{max}$&D$_{max}$\\ (kHz)&(ms)&(ms)&-&{(VIRGO)}&{(LIGO II)}\\ \hline
1.62&1.64&87.0&442&4.9&10.2\\ 1.83&1.25&27.0&155&6.4&13.5\\ 2.06&1.13&17.0&110&6.0&12.8\\
2.32&1.06&11.5&84&5.1&11.1\\ 2.72&1.00&8.4&72&3.6&5.7\\ \hline
\end{tabular}
\end{center}
\end{table}

After filtering the signal, the expected signal-to-noise ratio is

\begin{equation}
{(S/N)^2} = 4\int_0^{\infty}{{\mid \tilde h(\nu)\mid}\over{S_n(\nu)}}d\nu
\end{equation}

where $\tilde h(\nu)$ is the Fourier transform of the signal and
S$_n(\nu)$ is the noise power spectrum of the detector. Performing the required
calculations, the S/N ratio can be written as
\begin{equation}
{(S/N)^2} = {4\over 5}{h_0^2}\left({{\tau_{gw}}\over{S_n(\nu_{gw})}}\right){{Q^2}\over{1 + 4Q^2}}
\end{equation}
In the equation above, the angle average on the beam factors of the detector were already performed\cite{regimbau01}.

From eqs.( 4) and (11), once the energy and the S/N ratio are fixed, one can estimate
the maximum distance D$_{max}$ to the source probed by the detector. In the last two columns of
table 4 are given distances D$_{max}$ derived for a signal-to-noise ratio S/N = 2.0
and the sensitivity curve of the laser beam interferometers VIRGO and LIGO II.
In both cases, it was assumed that
neutron stars underwent the transition having a rotation period equal to the critical
value.

We emphasize again that our calculations are based on the assumption that the deconfinement
transition occurs in a dynamical timescale \cite{haensel82}. In the scenario developed
in ref.\cite{glen1}, a mixed quark-hadron phase appears and the complete deconfinement
of the core occurs according to a sequence of quasi-equilibrium states. The star
contracts slowly, decreasing its inertia moment and increasing the its angular velocity until
the final state be reached in a timescale of the order of $10^5$ yr \cite{glen1}. Clearly, in
this scenario no gravitational waves will be emitted and this could be a possibility to
discriminate both evolutionary paths.

\section{Conclusions}

The structure of pure hadronic configurations and that of hybrid stars having a quark
core, with the same baryonic number, were computed using a new equation of state\cite{marranghello01}.
The maximum stable mass for pure hadronic configurations satisfies the requirement of being higher
than the mass of Vela X-1 ( M = 1.86$\pm$0.16 M$_{\odot}$), determined recently\cite{barziv01}
from a new study of its orbital motion.

As one should expect,  the masses of quark cores increase with the baryon number of the
configuration, as well as the
difference between the binding energies between hybrid and single phase objects. On the
basis of this result and from the point of view of the GW emission, it would be natural to expect 
that phase transitions occurring in more massive stars would be more easy to detect.
In fact, this is not the case because, in the one hand the damping timescale due to
the emission of GWs is inversely proportional to the mass and hence massive
stars have low oscillation quality factors and, on the other hand the detectors
have sensitivities optimized at frequencies around 200 Hz. In comparison with
non-radial oscillations (for instance m= $\ell$ = 2 modes), these have comparable
frequencies but damping timescales
one order of magnitude higher than those resulting from radial modes coupled
to rotation. In this case, higher quality factors Q may be obtained but most of
the mechanical energy will be dissipated into heat. In order to be an efficient source of
GWs by the mechanism here considered, the star
must be a fast rotator, i.e., to have rotation periods of the order of few milliseconds.
Even in the most favorable case (model 2, corresponding to a baryonic mass equal to
1.32 M$_{\odot}$), if the rotation period is 4.0 ms, only 1\% of the available
energy will be emitted as GWs.

Inspection of table 4 indicates that the maximum distance probed by detectors of first
generation (VIRGO, LIGO I) is about 6.4 Mpc, well beyond M31, whereas
the second generation (LIGO II) will probably see phase transition events at distances
two times longer, but certainly not yet attaining the Virgo cluster. The small probed volume
and the rapid rotation required for this mechanism be efficient imply in a low event
rate, imposing  severe limitations on the detectability of such a signal.

\bibliography{guilhe}

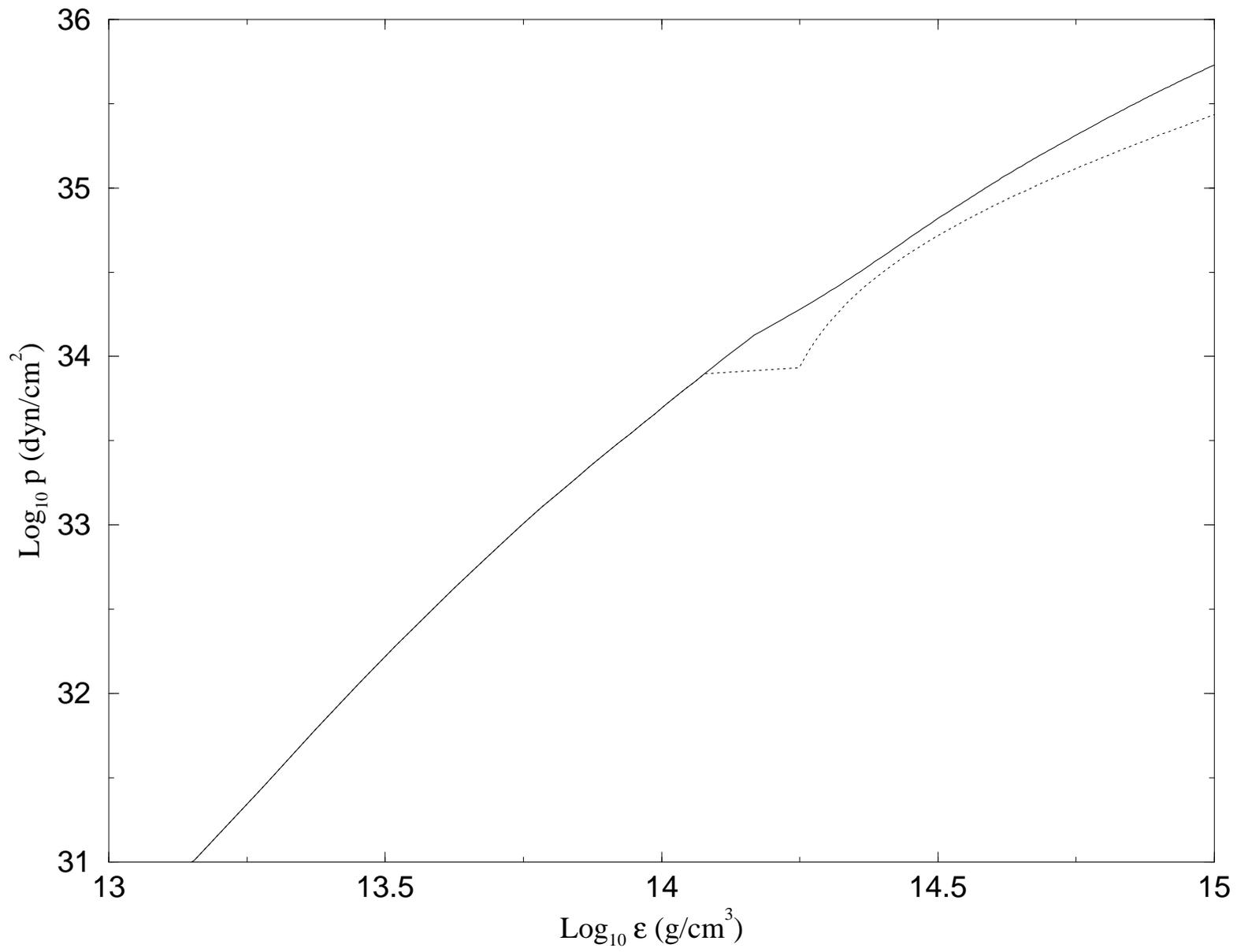
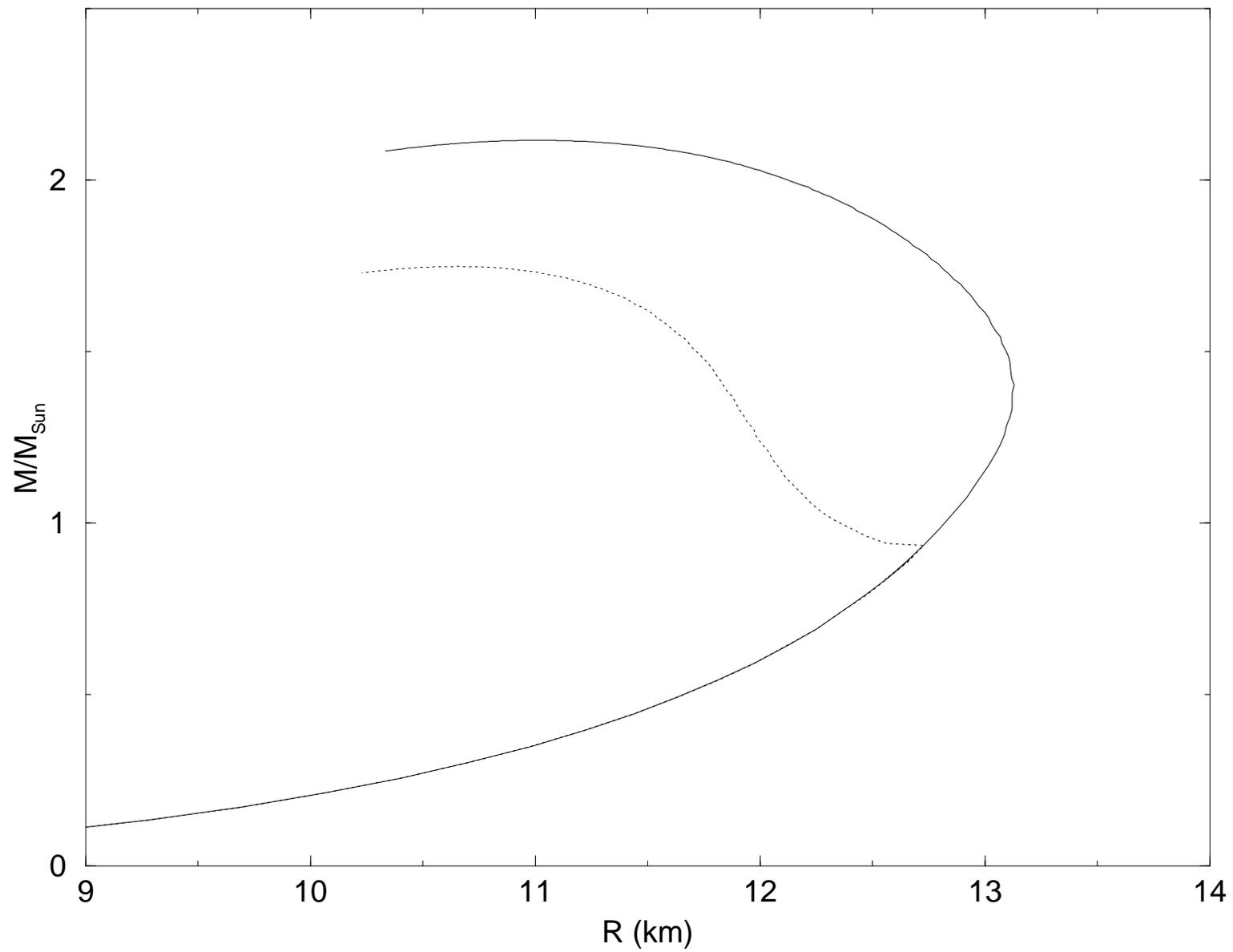
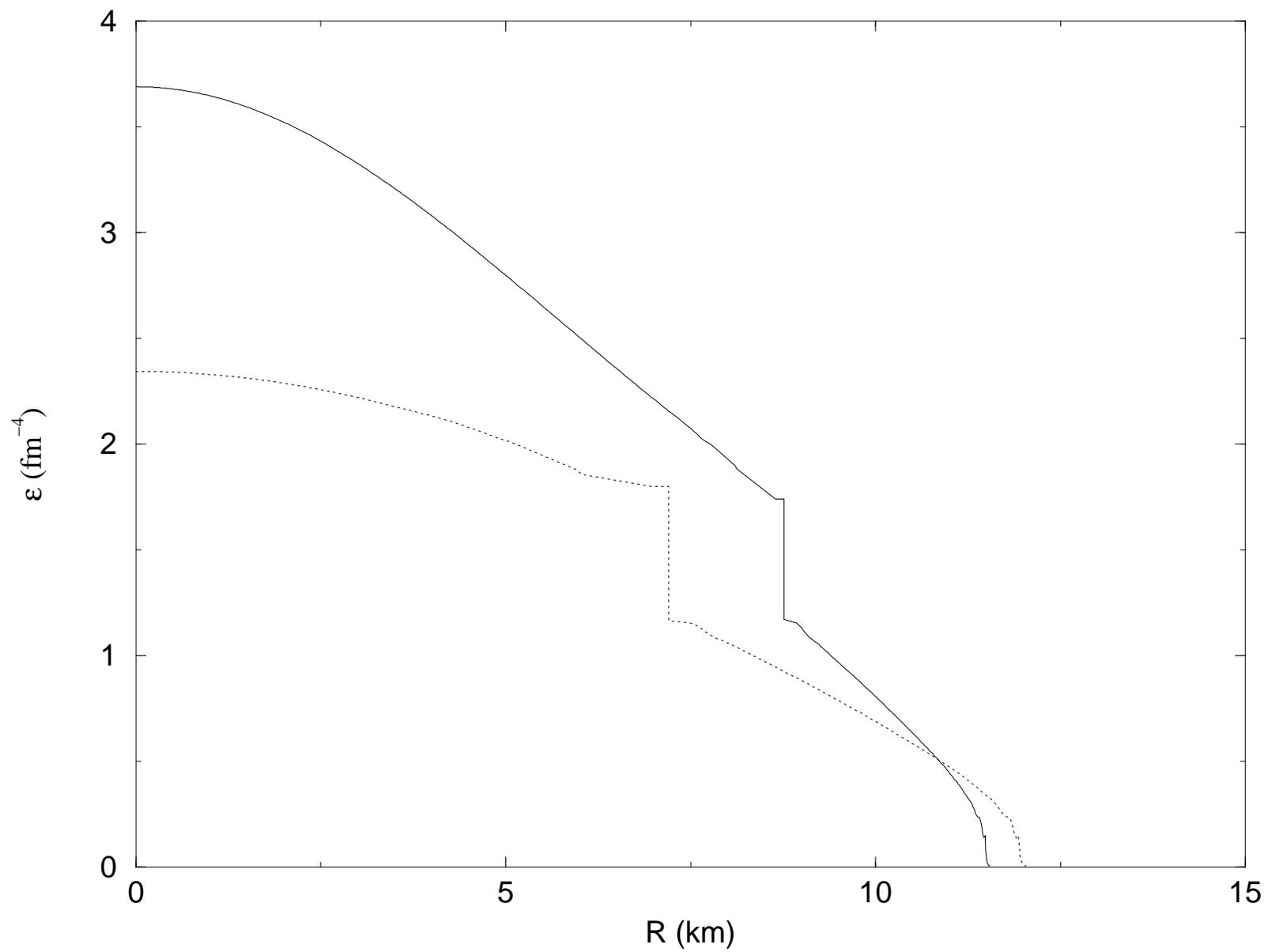
\end{document}